\documentclass[%
prb,
 amsmath,amssymb,
 superscriptaddress,
reprint,%
]{revtex4-2}

\usepackage{graphicx,hyperref}
\usepackage{dcolumn}
\usepackage{bm}

\usepackage[utf8]{inputenc}
\usepackage[T1]{fontenc}
\usepackage{amsmath}

\begin{document}

\preprint{}

\title[]{Unsupervised learning of non-Abelian multi-gap topological phases}

\author{Xiangxu He}%
\affiliation{Department of Physics and Institute for Advanced Study, The Hong Kong University of Science and Technology, Clear Water Bay, Kowloon, Hong Kong, China}

\author{Ruo-Yang Zhang}%
\affiliation{Department of Physics and Institute for Advanced Study, The Hong Kong University of Science and Technology, Clear Water Bay, Kowloon, Hong Kong, China}

\author{Xiaohan Cui}
\affiliation{Department of Physics and Institute for Advanced Study, The Hong Kong University of Science and Technology, Clear Water Bay, Kowloon, Hong Kong, China}

\author{Lei Zhang}%
\affiliation{State Key Laboratory of Quantum Optics Technologies and Devices, Institute of Laser Spectroscopy, Shanxi University, Taiyuan 030006, China}
\affiliation{Collaborative Innovation Center of Extreme Optics, Shanxi University, Taiyuan 030006, China}%

\author{C. T. Chan}
\email[Corresponding author: ]{phchan@ust.hk}
\affiliation{Department of Physics and Institute for Advanced Study, The Hong Kong University of Science and Technology, Clear Water Bay, Kowloon, Hong Kong, China}

\date{\today}

\begin{abstract}
Recent experiments have successfully realized multi-band non-Abelian topological insulators with parity-time symmetry. Their topological classification transcends the conventional ten-fold classification, necessitating the use of non-Abelian groups, manifesting novel properties that cannot be described using integer topological invariants.
The unique non-commutative multiplication of non-Abelian groups, along with the distinct topological classifications in the context of homotopy with or without a fixed base point, makes the identification of different non-Abelian topological phases more nuanced and challenging than in the Abelian case.
In this work, we present an unsupervised learning method based on diffusion maps to classify non-Abelian multi-gap topological phases. The automatic adiabatic pathfinding process in our method can correctly sort the samples in the same phase even though they are not connected by adiabatic paths in the sample set. Most importantly, our method can deduce the multiplication table of the non-Abelian topological charges in a data-driven manner without requiring \textit{a priori} knowledge. Additionally, our algorithm can provide the correct classifications for the samples within both the homotopy with and without a fixed base point. Our results provide insights for future studies on non-Abelian phase studies using machine learning approaches.
\end{abstract}

\maketitle
\section{Introduction} 
Non-Abelian states and statistics have extensively been studied\cite{nAs1,nAs2,nAs3,nAs4,nAs5}. Recently, the degenerate points in the parity-time ($PT$) symmetric multi-band systems have shown remarkable non-Abelian braiding structures, such as three-dimensional (3D) admissible nodal line configurations\cite{nAThe1,nAThe3,nAThe6,nAExp11,nAExp17} and the relations between monopole charge and its linking structure\cite{nAExp7,nAml1,nAml2,nAml3}. Research into non-Abelian topological physics has made exciting progress both theoretically\cite{nAThe1,nAThe2,nAThe3,nAThe4,nAThe5,nAThe6,nAThe7,nAThe8,nAThe9,nAThe10,nAThe11,nAThe12,nAThe13,nAThe14,nAThe15,nAThe16,nAThe17} and experimentally\cite{nAExp1,nAExp2,nAExp3,nAExp4,nAExp5,nAExp6,nAExp7,nAExp8,nAExp9,nAExp10,nAExp11,nAExp12,nAExp13,nAExp14,nAExp15,nAExp16,nAExp17}.
 Non-Abelian topological phases extend beyond the traditional tenfold classification\cite{tenfold1,tenfold2,tenfold3} and require characterization using non-Abelian groups. In Abelian systems, the 
 topological charges are integers obeying commutative additive relations and belonging to $Z$ or $Z_2$ Abelian groups.
However, non-Abelian topological charges exhibit non-commutative multiplication, such that $ij \neq ji$. This non-commutativity implies that exchanging the order of encircling two non-Abelian degeneracies in parameter space often results in different topological phases. Moreover, unlike Abelian topological phases whose classifications only rely on their symmetry classes, the classification of non-Abelian topological phases depends not only on symmetries but also on whether a fixed base point is present in the Hamiltonians\cite{nAThe6}. In the context of the homotopy with a fixed base point, the topological classification employs a specific non-Abelian group, whereas, in the absence of a fixed base point, the classification relies on the conjugacy classes of the non-Abelian group\cite{mermin}. Therefore, categorizing non-Abelian topological physical systems and investigating the properties of each classification is of great significance. Traditionally, classification can be achieved through the computation of topological charges. However, the complexity of the data prompts us to explore machine learning and data-driven approaches.

Machine learning has been extensively applied within the domain of physics\cite{mlph1,mlph2,mlph3,mlph4,mlph5}, spanning areas such as inverse design\cite{inv1,inv2,inv3,inv4,inv5} and the discovery of physical concepts\cite{pl1,pl2,pl3}. It leverages minimal physical knowledge, utilizing a data-driven approach to assist in resolving physical challenges. Many studies applying machine learning in physics rely on supervised learning\cite{su1,su2,su3,su4,su5}, which necessitates labeled data, and hence requires careful preparation of data sets. In contrast, unsupervised learning can discern data features without labels\cite{unp1,unp2}. This method is widely used in clustering\cite{uncl1,uncl2}, dimensionality reduction\cite{undd1,undd2}, and identifying association rules\cite{unar1}. Clustering algorithms achieved through unsupervised learning can effectively differentiate samples into distinct clusters while not assigning labels to the samples. Recently, some studies have explored the application of unsupervised learning in the classification of topological phases\cite{un1,un2,un3,un4,un5,un6,un7,un8,un9,un10,un11,un12,un13,un14,un15,un16,un17,un18,un19,un20}. However, the majority of research has focused on traditional Abelian topological phases. Moreover, most algorithms rely on existing adiabatic evolution paths within the sample set. When samples in the same phase are not all connected by adiabatic evolution paths in the sample set, they may be misclassified into different clusters. This issue is particularly pronounced in complex data sets, such as those involving non-Abelian phases.

In this work, we have employed an unsupervised learning method based on diffusion map\cite{dm1,dm2} to classify non-Abelian multi-gap topological phases. Our method incorporates a data-driven adiabatic pathfinding process before employing the diffusion map, which surpasses the constraints of earlier methods and enables effective clustering within the sample set without requiring any \textit{a priori} knowledge, even when certain samples within the same phases lack connectivity through adiabatic evolution paths. More importantly, our method can deduce the multiplication table of different topological phases without prior knowledge of the group structure, revealing the distinctive non-commutative multiplication relations between different non-Abelian topological charges. Consequently, our method can determine the topological classification of samples without computing the topological charge. Furthermore, our method is applicable to homotopy classifications both with and without a fixed base point, showing the different clustering results, which align well with topological theory. Our method broadens the application of machine learning within topological physical systems and lays the groundwork for future exploration of non-Abelian topological systems.

\section{Non-Abelian system}
\begin{figure*}    
\centering
\includegraphics[width=0.95\textwidth]{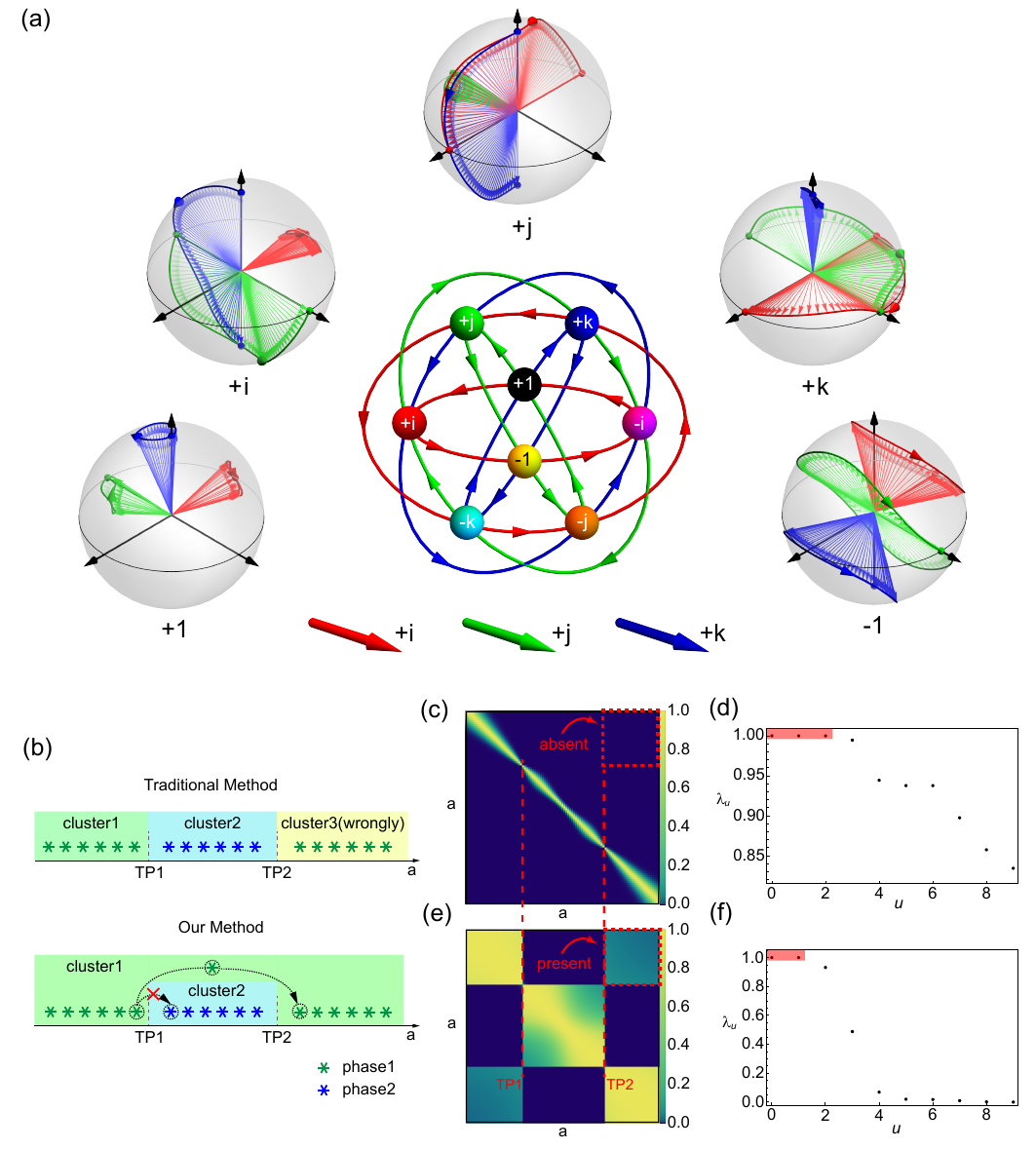}
\caption{(a)Schematic of Non-Abelian quaternion topological charges and Cayley graph of the quaternion group. The colored spheres in the central graph represent the elements of the quaternion group, while the colored arrows illustrate their mutual multiplication rules. Red, green, and blue arrows represent multiplication by $+i$, $+j$, and $+k$, respectively. In each surrounding sphere inset, the eigen-frame rotation characterizes a distinct topological quaternion phase. The red, green, and blue rotating arrows indicate the eigenvector evolution on the first, second, and third bands, respectively, as the momentum $k$ varies from 0 to $2\pi$.
(b)Comparison of the traditional unsupervised learning method and our improved method. Traditional clustering methods are easily interfered with by the topological transition points, which may lead to wrongly separating samples with the same topological charge to different clusters. Our method can find a temporary sample that can connect samples with the same charge but separated by the transition points, yielding the correct result.
(c)(e) Illustration of the local similarity matrices $K$ of the samples. (d)(f) The largest ten eigenvalues $\lambda_u$ of normalized local similarity matrices $T$. The number of eigenvalues $\lambda_u$ that satisfy $\lambda_u\approx1$ indicates the number of identified topological phases. With the traditional method, topological transition points separate the samples in the same topological phase, bringing the absence of the high similarity block in the normalized local similarity matrix $T$(shown by (c)), and incorrectly identifying three topological phases, as shown in (d). Our method can find the connection between the samples in the same topological phase, bringing the presence of the high similarity block (shown by (e)), and correctly identifying two topological phases(shown by (f)).
}

\label{FIG1}
\end{figure*}

We focus on the one-dimensional(1D) three-band $PT$-symmetric topological insulators with all bands being gapped\cite{nAExp1} as a prototypical example. With a well-chosen basis, the combination of the parity operator $P$ and the time-reversal operator $T$ is equivalent to the complex conjugate operator $K$. Therefore, the $PT$-symmetric Hamiltonian $H(k)$ satisfies $H(k)=H^*(k)$, where the asterisk symbol means complex conjugate and $k$ is the momentum. The PT symmetry guarantees that the Hamiltonian remains real in the whole $k$-space. A 1D three-band $PT$-symmetric Hamiltonian $H(k)$ with two complete bandgaps can be written as
\begin{equation}
H(k) | \psi_k ^n\rangle=\varepsilon_k^n| \psi_k^n \rangle,\ n\in\{1,2,3\}.
\end{equation}
Here, $\varepsilon_k^n$ and $| \psi_k^n \rangle$ are the $n$-th eigenvalue and the normalized eigenvector. Notice that $| \psi_k^n \rangle$ is a real 3D vector. Due to the complete bandgaps, the eigenvalues $\varepsilon_k^n$ satisfies
\begin{equation}\varepsilon_k^i\neq \varepsilon_k^j, \ i\neq j, \ i,j \in \{1,2,3\},\end{equation}
for all $k$ in the momentum space. Without loss of generality, we consider the momentum in the first Brillouin zone, namely, $k\in[0,2\pi]$. According to past research\cite{mermin}, the topological classification of such 1D Hamiltonians depends on the fundamental homotopy group of the order-parameter space. In the traditional Abelian system, the order-parameter is often set to be the eigenvector of a single band. In contrast, the order-parameter space of the non-Abelian system is the eigenvector frame consisting of multiple eigenvectors. The rotation of the eigenvector frame across the first Brillouin zone determines the topological phase of the system. The corresponding order-parameter space of the 1D three-band Hamiltonian is $M=\frac{O(3)}{O(1)^3}$, where $O(N)$ is the orthogonal group. The $O(1)$ group in the denominator represents the equivalence of eigenvectors with different signs. The fundamental homotopy group of the system can be expressed as\cite{mermin} 
\begin{equation}\pi_1(M)=Q= \{ 1, \pm i, \pm j, \pm k, -1 \},\end{equation}
which is the non-Abelian quaternion group with the multiplication relationship of three anti-commuting imaginary units $ij=k,jk=i,ki=j,i^2=j^2=k^2=-1$. Fig. \ref{FIG1}(a) illustrates the elements of the quaternion group and their mutual multiplication (depicted by colored arrows). 

To show the frame rotation properties more explicitly, we can express the three-band Hamiltonian as
\begin{equation}H(k)=R(k)\text{diag($\varepsilon_k^1,\varepsilon_k^2,\varepsilon_k^3$)}R(k)^\intercal,\end{equation}
with the eigenvector frame $R(k)=(|\psi_k^1\rangle,|\psi_k^2\rangle,|\psi_k^3\rangle)=\text{exp}(\phi_k \mathbf{n}_k\cdot\mathbf{L})$, where $(\mathbf{L}_i)_{jk}=-\epsilon_{ijk}$ is the fully antisymmetric tensor, $\phi_k$ and $\mathbf{n}_k$ are the rotation angle and rotation axis, respectively. When $k$ varies from $0$ to $2\pi$, the eigenvector frame $R(k)$ rotates in the order-parameter space. In the insets of Fig. 1(a), we display the tips of the eigenvectors and show how the eigenvector of each band (indicated by the color) rotates as $k$ varies from $0$ to $2\pi$. As the surrounding sphere insets in Fig. \ref{FIG1}(a) show, several types of rotation patterns are possible, corresponding to different quaternion topological charges that describe 3D rotation. Charges $+i$, $+j$, $+k$ indicate that when $k$ varies from $0$ to $2\pi$, one eigenvector will return to the initial position, while the others settle at the antipodal points to their initial position on the sphere. Charge $+1$ indicates the situation of all eigenvectors also return to their initial positions, and the rotation trajectory of the eigenvector frame can contract to the unit vector frame, namely, $((1,0,0)^\intercal, (0,1,0)^\intercal,(0,0,1)^\intercal)$. In the case of charge $-1$, all the eigenvectors go back to the initial position, too. However, the trajectory of the rotation of the eigenvector frame cannot contract to the unit vector frame. 

Charge $-i$($-j, -k$) is almost the same with charge $+i$($+j, +k$), but they take opposite eigenframe rotation directions. It should be noticed that the difference between charge $+i(+j, +k)$ and its opposite charge only exists in the homotopy with a fixed point\cite{nAThe6}. Physically, a base point for a group of Hamiltonians means that all Hamiltonians share a common matrix form $H(k_0)$ at a certain $k_0$. If one refers to the homotopy without a fixed base point (free homotopy), charge$ +i(+j, +k)$ is equivalent to charge $-i(-j,-k)$ resulting in the topological classification with the conjugacy classes of quaternion group: $\{+1\}, \{\pm j\}, \{\pm j\}, \{\pm k\}, \{-1\}$.  The free-homotopy classification is often regarded as a more natural characterization of non-Abelian topological phases since actual physical systems can only guarantee to satisfy certain symmetries but not have a common base point.

Therefore, 1D three-band $PT$-symmetric topological insulators can be topologically classified using the quaternion group that describes 3D rotation. Notice that the loops surrounding the nodal lines in a 3D three-band gapless system can be classified in the same manner\cite{mermin}. Our mission is to cluster samples in such a system using machine learning.

\section{Diffusion map}
To introduce the diffusion map, we consider a sample set $V = \{v_1,v_2,...,v_m\}$, where $m$ is the number of samples. In our case, $V$ refers to the dataset of the eigenvector frames $R(k)$. Here, $k$ runs over the first Brillouin zone, taking the value from $0$ to $2\pi$. The local similarity between the samples $v_i$ and $v_j$ can be expressed in a Gaussian form
\begin{equation}K_{i,j}=\text{exp}[-\frac {d(v_i,v_j)} {\zeta }].\end{equation}
Here,  $\zeta$ is the clustering parameter, and we set $\zeta=0.01$. $d(v_i,v_j)$ denotes the distance between the data points $v_i,v_j$. In general, this distance could be defined in various ways. Here, we defined it as
\begin{equation}d(v_i,v_j)=1-\frac{1}{N_kN_{band}}
\sum_k\sum_n \left| \langle \psi_{i,k}^n \mid \psi_{j,k}^n\rangle \right|^2,\end{equation}
with eigenvectors $| \psi_{i,k}^n \rangle$ satisfying $H_{i,k}| \psi_{i,k}^n \rangle=\varepsilon_{i,k}^n | \psi_{i,k}^n \rangle$.
Here, $N_{band}$ is the number of bands and $N_k$ is the total number of discrete $k$ we sample in $[0,2\pi]$. In our case, $N_{band}=3$. To balance between computational speed and accuracy, we took $N_k=51$ discrete $k$ points evenly distributed from $0$ to $2\pi$.

A Markovian random walk is conducted on the dataset $V$, while the normalized local similarity, or namely, the one-step diffusion matrix $T_{i,j}$ serves as the one-step diffusion probability from the sample $v_i$ to the sample $v_j$
\begin{equation}T_{i,j}=\frac {K_{i,j}} {\sum_j K_{i,j}}.\end{equation}
The diffusion distance between samples $i$ and $j$ after $2t$ steps can be described by \begin{equation}D_{2t}(i,j)
=\sum_o \frac{\left|(T_{i,o})^t-(T_{j,o})^t\right|^2}{\sum_{o'}K_{o',o}}
=\sum_{u=1}^{N-1}\lambda^{2t}_u \left|\phi_{u,i}-\phi_{u,j}\right|^2,\end{equation}
where $\lambda_u$ and $\phi_u$ are the  $u$th eigenvalues and eigenvectors of matrix $T$, i.e., $T\phi_u=\lambda_u\phi_u$, with the order $\lambda_0=1\ge\lambda_1\ge\lambda_2\ge...\ge\lambda_{m-1}$. After a long time diffusion $t\rightarrow\infty$, only the terms $\phi_u$ with largest $\lambda_u\approx1$ will dominate. We can use these few components $\phi_u$ to represent the samples so as to reduce the dimensions. Additionally, the number of dominated terms determines the number of clusters.

\section{Data-driven adiabatic pathfinding method}
The traditional diffusion map process has been described in the previous paragraph. However, this conventional method implicitly assumes that there are sufficient adiabatic evolution paths present in the sample set, allowing samples in the same topological phase to diffuse through Markov random walks. If such paths do not exist within the sample set, samples in the same topological phase may be divided into multiple clusters, with each cluster containing only the connected subset of samples, as illustrated in the upper panel in Fig. \ref{FIG1}(b). This issue is particularly pronounced in our non-Abelian system due to the complexity of the dataset. We generated the data by varying a parameter in a three-band non-Abelian Hamiltonian (see Appendix I). Fig.~\ref{FIG1}(b) shows that changes in parameters caused the samples to continuously pass through two topological phase transition points. According to the knowledge of topological theory, we found that the left and right sections belong to the same topological phase (shown as green asterisks), while the middle section belongs to a different topological phase (shown as blue asterisks). However, there are no adiabatic evolution paths connecting the samples in the left section and those in the right section, leading to an incorrect classification of the sample set into three categories by traditional methods. The heat map of the matrix $K_{i,j}$ (Fig. \ref{FIG1}(c)), where the color intensity represents the similarity between the sample $i$ and $j$, indicates a lack of connectivity between the samples in the left and right sections, as the absent high similarity block shows (symboled by the red dashed square in Fig. \ref{FIG1}(c)). The three largest eigenvalues $\lambda_u$ of the one-step diffusion matrix $T$ approach to 1 (Fig. \ref{FIG1}(d)), which indicates that two topological phases are wrongly sorted into three clusters.

\begin{figure*}
\centering
\includegraphics[width=0.95\textwidth]{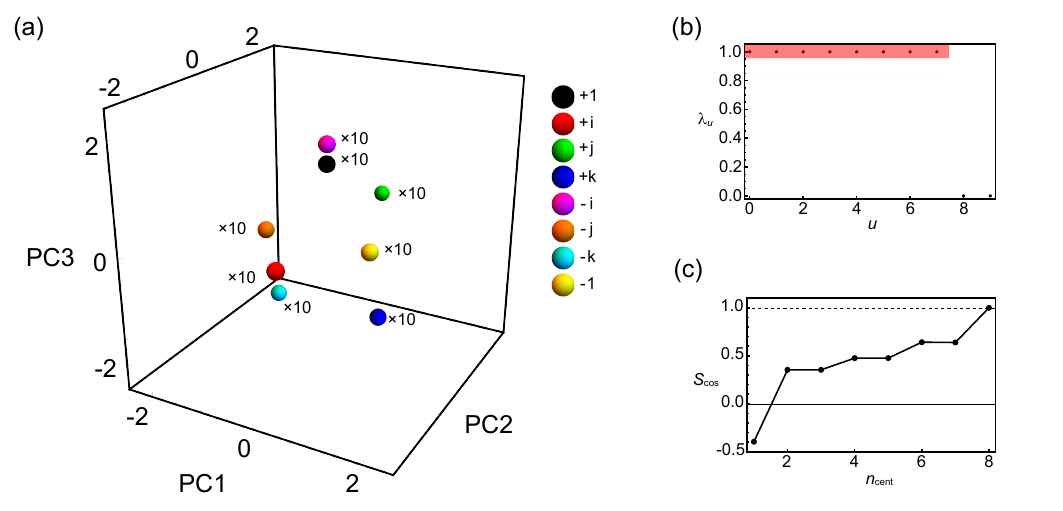}
\caption{(a)Scatter diagram of samples in the space of three principal components (PC1, PC2, PC3), which are linear combinations of the eigenvectors of the one-step diffusion matrix $T$. Samples are clustered into eight topological phases, with ten samples clustered at each center, nearly at the same position. (b)Eigenvalues $\lambda_u$ of the one-step diffusion matrix $T$. The eight largest eigenvalues $\lambda_{0,1,2,3,4,5,6,7}$, correspond to the eight topological phases. (c)The result of the k-means algorithm applied to the clustered sample set. The horizontal axis shows the ordinal of centroids $n_{\text{cent}}$, while the vertical axis represents the cosine similarity $S_{\text{cos}}$. When the assumed number of centroids $n_{\text{cent}}$ reaches $8$, the cosine similarity approaches its maximum value of $1$. It indicates that the samples have been effectively divided into eight distinct topological phases.
}
\label{FIG2}
\end{figure*}

To address this issue, we have incorporated a data-driven adiabatic pathfinding process. This approach automatically identifies potential adiabatic evolution paths between the samples to connect them, effectively connecting previously segregated samples by generating intermediate samples, as illustrated schematically in the lower panel of Fig. \ref{FIG1}(b). The process enables the sample to try to adiabatically transform to another sample. If the initial sample and the target sample belong to the same phase, an adiabatic evolution path can be found to connect each other. If they are from distinct phases, there must be no adiabatic path connecting them, and the transforming process must be interrupted by the phase transition points. To make it clearer, there will be an intermediate state too close to the singularity and cannot meet the continuity criteria (see Appendix II for details). Equipped with the adiabatic pathfinding process, we did the clustering to the same dataset, and the result is shown in Fig. \ref{FIG1}(e)(f). We can see the high similarity between the left section and the right section (the green block labeled by the red dashed square in Fig. \ref{FIG1}(e)) appears, indicating the adiabatic paths connecting these samples are detected. The eigenvalue graph of the one-step diffusion matrix also shows the two largest eigenvalues satisfying $\lambda_u\approx1$, which align with the two phases in the dataset.

\section{Case of Homotopy with a Fixed Base Point}
In the following, we present the result of clustering a sample set consisting of three-band non-Abelian $PT$-symmetric Hamiltonians. The sample set was generated according to the procedure described in Appendix I. It is important to note that our samples share a fixed base point, which means that when $k=0$, all samples assume the same Hamiltonian $H(k=0)$.

The results are illustrated in Fig. \ref{FIG2}. The scatter diagram(Fig. \ref{FIG2}(a)) demonstrates that the samples are clustered into eight different topological phases in the space spanned by the eigenvectors $\phi_u$ of the one-step diffusion matrix $T$. Samples within the same topological phase are clustered at nearly the same position, and every clustering center in Fig. \ref{FIG2}(a) includes ten samples. Fig. \ref{FIG2}(b) displays the corresponding ten largest eigenvalues of $T$: $\lambda_{0,1,\dots,9}$. We can see that eight of the largest $\lambda_u$ satisfy $\lambda_{0,1,\cdots7}\approx1$, aligning with the eight topological phases. To further confirm that the samples have been clustered into eight phases, we employed the k-means algorithm\cite{kmeans} to determine the number of k-means centroids for the dataset. In applying the k-means algorithm, we preset the number of centroids, after which the algorithm automatically identified the positions of centroids as clustering centers for the samples. Finally, we can use cosine similarity to evaluate whether the preset number of centroids is suitable for the data. When the cosine similarity $S_{cos}$ approaches $1$ for a specific number of centroids, each sample is nearly at the same position with one clustering center, indicating that we get the optimal number of centroids. As shown in Fig. \ref{FIG2}(c), the cosine similarity $S_{cos}$ reaches $1$ only when the number of centroids increases to $8$. Thus, the k-means algorithm also confirms that the sample set has been clustered into eight topological phases. These results align with the theoretical calculation of the topological charges, although our method does not involve such calculations.

\begin{figure*}
\centering
\includegraphics[width=0.95\textwidth]{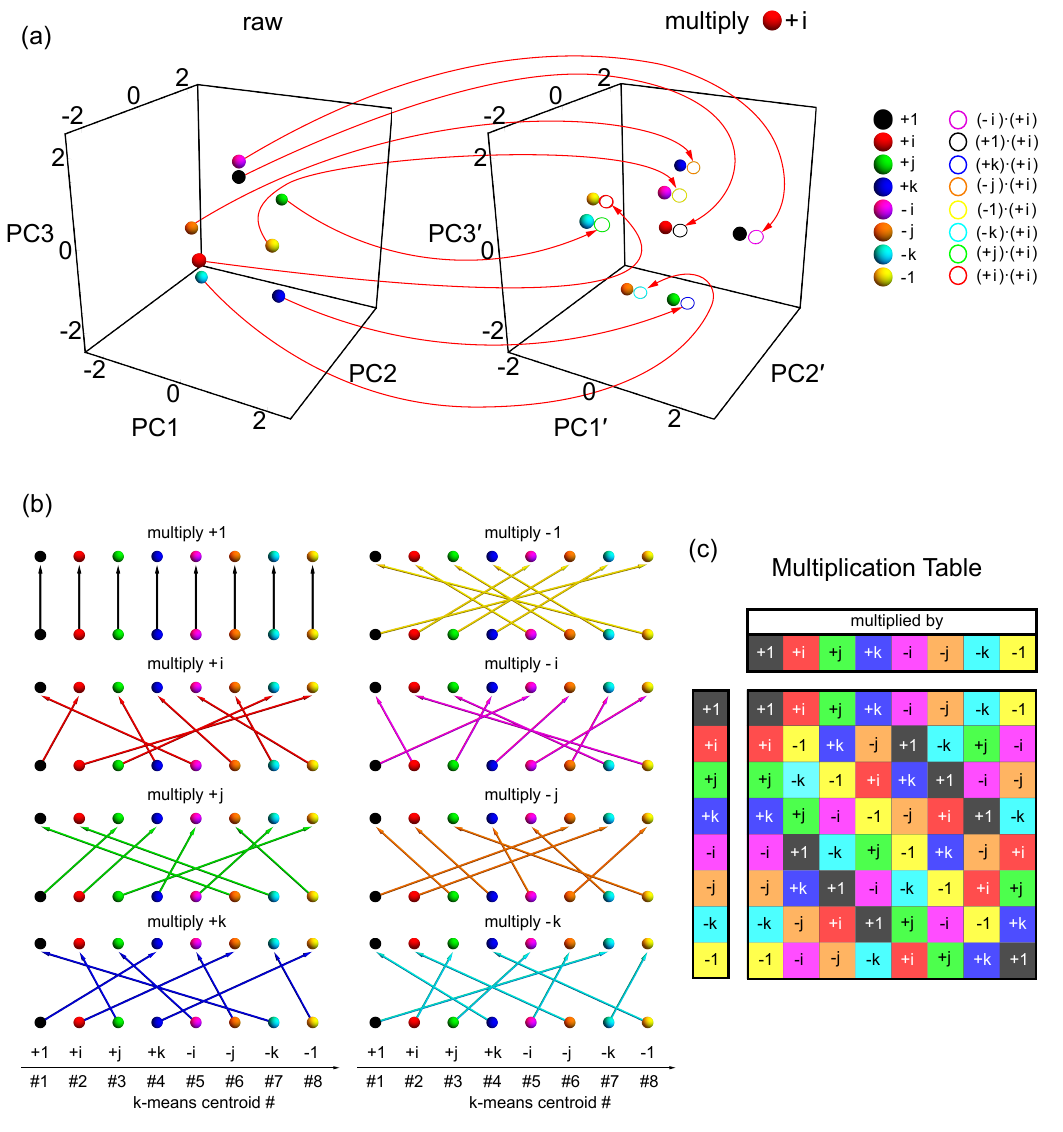}
\caption{(a)Schematic of learning the multiplication rule of $Q*(+i)\rightarrow Q$ (second column in the (c) multiplication table). The left panel displays the scatter diagram of the clustered original dataset $V$, while the right panel shows the scatter diagram of the clustered mixed dataset $V_m$, where the solid spheres represent the dataset $V$ and the hollow circles denote the dataset $V_p$. The red curve arrows represent the correspondence of the clusters before and after the multiplication. (b) Unsupervised learning results of all isomorphism $Q*g\rightarrow Q$ with $g\in Q$ (each column in the multiplication table). It shows the multiplication correspondence of different topological phases. (c)Multiplication table of different topological phases. It has a one-to-one correspondence with the multiplication table of the quaternion group.
}
\label{FIG3}
\end{figure*}

\section{Multiplication Table}
In this section, we offer a detailed description of how our method derives the multiplication table of different topological phases without the knowledge of topological theory, obtaining the algebraic structure of the quaternion group without requiring \textit{a priori} knowledge. 

Non-Abelian phases obey a remarkably unique non-commutative multiplication rule, as illustrated by the colored arrows in Fig. \ref{FIG1}. To explore the multiplication between non-Abelian phases, we introduce the ``multiplication'' of two given Hamiltonians $H_1(k),H_2(k)$ sharing a common base point $H_1(k=0)=H_2(k=0)$ as a new Hamiltonian $H_{12}=H_1\circ H_2$. Here, ``$\circ$'' represents a ``multiplication operator'' between two mappings defined as (should not be confused with the matrix product)
\begin{equation}H_{12}(k)=(H_1\circ H_2)(k)\equiv\left\{ \begin{array}{ll}
	H_1(2k),&k\in[0,\pi]\\
	H_2(2k-2\pi),&k\in[\pi,2\pi]\\
\end{array} \right. \end{equation}
The non-commutative nature of non-Abelian group multiplication is crucial, as changing the order of multiplication generally changes the result, i.e. $H_{12}(k)\neq H_{21}(k)$ in general cases. When generating the samples, we use the operation on the loops (See Appendix I). In this case, the multiplication of two samples is illustrated as Fig. \ref{FIG6}.

Fig. \ref{FIG3}(a) demonstrates our method for identifying the topological charge of the multiplication between two topological phases. With the defined multiplication, we proceed to operate on the original sample set $V$, previously clustered in the former step (left panel of Fig. \ref{FIG3}(a)). We randomly selected one sample $v_j$ from a cluster. Here, we took the sample from the cluster with non-Abelian topological charge $+i$ without loss of generality. We then multiplied each sample in the original sample set $V$ by sample $v_j$ forming a product dataset $V_p=\{v_i\circ v_j|v_i\in V\}$. Subsequently, we combined the original dataset $V$ with the product dataset $V_p$ to get a mixed dataset $V_m=V\cup V_p$. We used the mixed dataset as the input and performed clustering using the aforementioned method. The result is shown in Fig. \ref{FIG3}(a). The left panel of Fig. \ref{FIG3}(a) is the scatter diagram for the original dataset $V$, while the right panel shows the scatter diagram for the mixed dataset $V_m$. In the right panel, the solid spheres represent the samples from the original dataset $V$, whereas the hollow circles denote the product dataset $V_p$. Each pair of adjacent solid sphere and hollow circle represents a single resultant cluster in the mixed dataset. The red curve arrows indicate the correspondence of the clusters before and after the multiplication. For example, the samples plotted as the green spheres ($+j$) in the left panel are mapped to the green hollow circles in the right panel after multiplication $(+j)\cdot(+i)$, which is adjacent to the solid cyan sphere $(-k)$ indicating the multiplication result $-k=(+j)\cdot(+i)$. Remarkably, the mixed dataset is also clustered into eight topological phases, which demonstrates the closure of the multiplication operation. 
Therefore, our data-driven method successfully derives the multiplication table of the topological phases. 

We selected samples from all eight clusters of the original dataset $V$, and applied the processes described above. Finally, we obtained eight mixed data sets $V_{m1,m2,\dots,m8}$ along with their corresponding clustering results. 
For clarity, we applied k-means algorithms to the clustering results, and we focused on the one-to-one correspondence of the clusters before and after the multiplication mentioned above. This analysis yielded the content presented in Fig. \ref{FIG3}(b). The samples from the mixed dataset $V_{mi}, i=1,2,\dots,8$ are clustered into eight topological phases, allowing us to assign a k-means centroid ordinal for each cluster. The horizontal axes in Fig. \ref{FIG3}(b) represent the ordinals of the k-means centroids. In each panel of Fig.~\ref{FIG3}(b), the lower solid spheres ($V_i$) indicate the cluster positions before multiplication, while the upper solid spheres ($V_{mi}$) are the positions after multiplied by a specific topological charge (indicated by the color of the arrows). These arrows illustrate the one-to-one correspondence between the topological phases before and after multiplications.

By summarizing all multiplication results, we derived the multiplication table of the topological phases, depicted in Fig. \ref{FIG3}(c). For example, the multiplication result of $V\cdot (+i)$  is shown in the second column of the square table in Fig. \ref{FIG3}(c). With the multiplication table, we can easily identify that the non-commutative multiplication relations between the samples obey the algebraic structure of the non-Abelian quaternion group.

Importantly, the aforementioned processes do not require labeling or prior knowledge of the exact topological charge of each cluster. Once the multiplication table of the topological phases is obtained, as shown in \ref{FIG3}(c), we can compare it with the quaternion group's multiplication table, identifying the one-to-one correspondence between our sample's topological phases and the elements of the non-Abelian quaternion group. Consequently, we can assign a topological charge in the quaternion group to each sample without the need for traditional calculations of topological charge. This achievement represents the "labeling" of the unsupervised learning sample set in a data-driven manner.

\section{Case of Free Homotopy}
Next, we consider the topological classification within the context of free homotopy. As previously noted, actual physical systems can only be ensured to satisfy certain symmetries, but hard to maintain a common base point during the variation of parameters. Thus, it is more natural to discuss the topological classification in the context of the homotopy without a fixed base point, which is also called the free homotopy. In free homotopy, samples before and after a conjugate operation are considered equivalent. For the 1D three-band $PT$-symmetric topological insulators, the inequivalent topological phases in the sense of free homotopy are classified by the conjugacy classes of the quaternion group: $\{+1\}$, $\{\pm i\}$, $\{\pm j\}$, $\{\pm k\}$, $\{-1\}$\cite{nAThe6}. Thus, the difference between charges with opposite signs vanishes, reducing the number of topological phases to five.

\begin{figure*}
\centering
\includegraphics[width=0.95\textwidth]{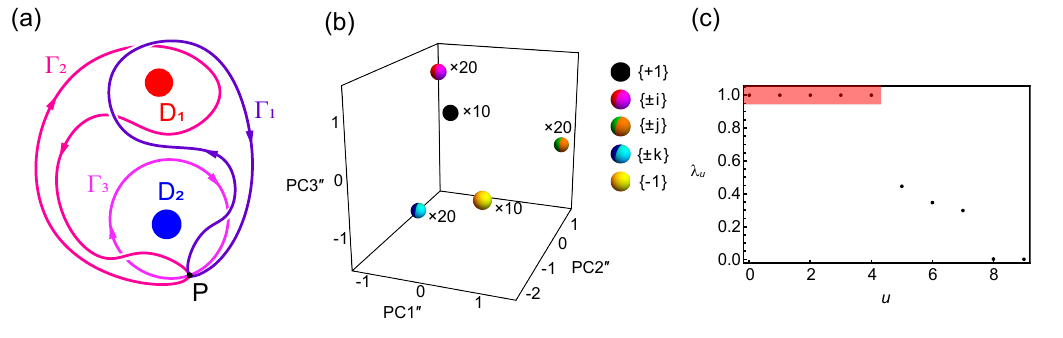}
\caption{(a) Comparison of homotopic equivalence relations of loops with and without a fixed base point. $\Gamma_1,\Gamma_2$ and $\Gamma_3$ are loops starting from the fixed base point $P$. D$_1$ and D$_2$ are different degeneracies in the plane. $\Gamma_1$ and $\Gamma_2$ are topologically inequivalent under based homotopy (with a base point). However, they are equivalent under free homotopy (without a base point), because $\Gamma_2=\Gamma_3\Gamma _1\Gamma_3^{-1}$.  (b) Samples are clustered into five topological phases. Notice that samples with two distinct topological charges $+i$ and $-i$ are merged into a single cluster under free homotopy. Similar merging occurs for the pair $+j$ and $-j$ ($+k$ and $-k$). (c) Five largest eigenvalues $\lambda_{0,1,2,3,4}$ of the one-step diffusion matrix $T$ approach to 1, indicate five distinct topological phases under free homotopy. 
}
\label{FIG4}
\end{figure*}

To intuitively illustrate the conjugate transformation and its implications, we show a loop model in Fig. \ref{FIG4}(a). $\Gamma_1$, $\Gamma_2$, and $\Gamma_3$, all sharing a common point $P$, are three loops on a plane containing two degeneracies $\text{D}_1$ and $\text{D}_2$ (red and blue circles) formed by bands 1,2, and bands 2,3, respectively. Both $\Gamma_1$ and $\Gamma_2$ encircle the degeneracy $\text{D}_1$, while $\Gamma_3$ encircles the degeneracy $\text{D}_2$. $\Gamma_1$ is a loop that traverses the "left side" of the degeneracy $\text{D}_2$, whereas $\Gamma_2$ traverses the "right side". When considering homotopy with a fixed base point $P$, $\Gamma_1,\Gamma_2$ are not topologically equivalent,  since they cannot be continuously transformed into each other without leaving the point $P$. However, the two loops can be continuously transformed into each other in the absence of a fixed base point. This topological equivalence in the free homotopy sense manifests as the two loops are connected to each other by a conjugation operation: $\Gamma_2 = \Gamma_3 \Gamma_1\Gamma_3^{-1}$. 

Similarly, we employ the definition of multiplication of the samples to define the conjugate operation. When we use the sample $H_1$ to perform the conjugate operation to the sample $H_2$, we have
\begin{equation}H_c=H_1\circ H_2 \circ \bar{H}_1\equiv\left\{ \begin{array}{ll}
	H_1(4k),&k\in[0,\pi/2]\\
	H_2(2k-\pi),&k\in[\pi/2,3\pi/2]\\
        H_1(8\pi-4k),&k\in[3\pi/2,2\pi]\\
\end{array} \right. \end{equation}
Here, $\bar{H}_1(k)=H_1(-k)$. To obtain results in the context of the free homotopy, we selected a sample to perform the conjugate operation on all the samples from the original sample set, treating the samples before and after the conjugate operation as equivalent. 

Our results are presented in Fig. \ref{FIG4}(b)(c). The scatter diagram in Fig. \ref{FIG4}(b) shows that the sample set is clustered into five topological phases. Notably, examples with topological charges of $+i$ and $-i$ are clustered at nearly the same location. Similarly, $+j,-j$  and $+k,-k$  also merge into the same clusters, respectively. The samples with charge $+1$($-1$) do not merge with other samples with different quaternion charges. Fig. \ref{FIG4}(c) shows the five largest eigenvalues $\lambda_{0,1,2,3,4}\approx1$ of the one-step diffusion matrix $T$, corresponding to the results of the five topological phases in the absence of a fixed base point.

\section{Discussion and conclusion}
In summary, we have introduced a method that employs unsupervised learning to classify non-Abelian topological multi-gap phases. Our approach incorporates a data-driven pathfinding process, enabling it to connect samples within the same topological phase that are otherwise disconnected in the sample set, thus achieving accurate non-Abelian topological classification without any \textit{a priori} knowledge. More importantly, our algorithm derived the multiplication table of the non-Abelian topological phases without prior knowledge of the group structure. Therefore, our method can identify the group algebraic structure of the quaternion group through the multiplication relationship between the samples in a data-driven manner. With the multiplication properties of the samples, we can assign the topological charge to every sample, realizing the labeling of the sample set without calculating the topological charge in the traditional way. Additionally, we performed clustering of the sample sets both in the presence and absence of a fixed base point, revealing two distinct topological classification results: the quaternion group itself and its conjugacy classes. This result not only demonstrates the generality of our method but also highlights the role of the base point in non-Abelian topological phase classifications.

Notably, recent studies have utilized unsupervised learning to classify non-Abelian eigenvalue braidings in non-Hermitian systems\cite{un19,un20}. However, the algorithms presented in these studies mainly rely on identifying phase transition points, and neither identify the non-Abelian multiplication rules nor discuss the role of the base point. In contrast, all these issues crucial for non-Abelian topological phase classifications have been resolved in our work. Our research broadens the application of machine learning in topological physics and paves the way for further exploration of non-Abelian multi-gap systems.

\begin{figure*}
\centering
\includegraphics[width=0.95\textwidth]{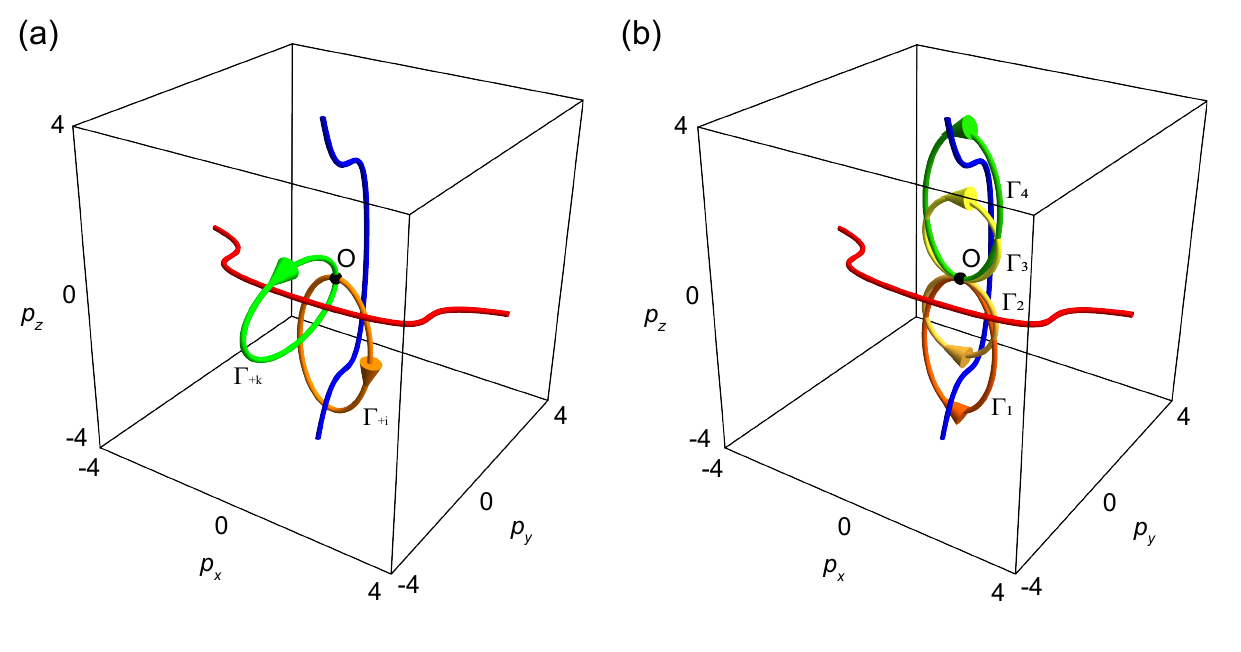}
\caption{(a)The nodal line configuration of the prototype Hamiltonian $\mathcal{H}(\mathbf{l})$ and the closed loop we use to generate the 1D Hamiltonians $H(k)$. The red(blue) line represents the degeneracy between band 1,2 (band 2,3). A green loop encircles the nodal line, with a green arrow illustrating the loop's orientation, while the black point $O$ marks the base point of the loop. (b) The loop series we use to generate samples for Fig. \ref{FIG1}(b-f). By fixing the other parameters of the elliptical loop (with a semi-minor axis of $1.0$, starting point at the origin, and constrained within the $k_y=0$ plane) while varying the semi-major axis from $-\frac{10} {3}$ to $\frac{10} {3}$ (where the sign indicates the direction of the axis), we obtain a series of loops. The orange, light orange, yellow, and green loops correspond to semi-major axes of $-1.8$, $-1.0,$ $1.0$, and $1.8$, respectively. Both the orange and green loops encircle the blue nodal line, whereas the light orange and yellow loops do not encircle any nodal line. 
}
\label{FIG5}
\end{figure*}

\begin{acknowledgments}
We would like to thank Prof. Biao Yang for the useful discussions. This work is supported by Hong Kong RGC AoE/P-502/20 and National Natural Science Foundation of China Grant No. 12474047.
\end{acknowledgments}

\section*{Appendix}

\subsection{Generation of the Dataset.}
As previously mentioned, the topological classification of 1D three-band $PT$-symmetric insulator is the same as the classification of nodal lines in the 3D three-band $PT$-symmetric gapless Hamiltonians based on homotopy. Therefore, we generate the 1D Hamiltonians from the loops in an auxiliary 3D space spanned by $\mathbf{p}=(p_x,p_y,p_z)$ as our dataset.

We consider a 3D $PT$-symmetric prototype Hamiltonian
\begin{equation}
\begin{aligned}
 &\mathcal{H}(\mathbf{p}) 
=\left(\begin{array}{lll}
p_y^3 & t p_x & t p_z \\
t p_x & -p_y+(b+p_y^2) & c p_x p_z \\
t p_z & c p_x p_z & -p_y-(b+p_y^2)
\end{array}\right)
\end{aligned},
\end{equation}
In this Hamiltonian, we set $t=1,c=\frac{1}{4},b=2$. This system features two continuous nodal lines formed by the adjacent bands, respectively. As demonstrated in Fig. \ref{FIG5}(a), the red and the blue lines represent the nodal lines derived from the degeneracies of bands 1,2 and bands 2,3, respectively.

A closed loop in the auxiliary 3D space can be described by a function $\Gamma(k)$:
\begin{equation}
\Gamma:[0,2\pi] \rightarrow \boldsymbol{R}^3,\quad \Gamma(0)=\Gamma(2\pi). \end{equation}
Whenever the loop does not intersect the nodal lines (such as the green loop in Fig. \ref{FIG5}(a)), we obtain a 1D gapped Hamiltonian from the loop: $H(k)=\mathcal{H}(\Gamma(k))$. Here, the parameter $k$ corresponds to the Bloch wavenumber of the generated 1D Hamiltonian. As depicted in Fig. \ref{FIG5}(a), the loops are marked in the green curve and orange curve, and the arrow represents the orientation of the loop when $k$ increases. When the loop encircles a nodal line, the eigenvector frame of the 1D gaped Hamiltonian $H(k)$ will show the non-trivial rotation in the order-parameter space, resulting in non-trivial topological characteristics. In Fig. \ref{FIG5}(a), the green loop $\Gamma_{+k}$ encircles the red nodal line, resulting in the non-Abelian phase $+k$, while the orange loop $\Gamma_{+i}$ encircles the blue nodal line, resulting in the non-Abelian phase $+i$. If it does not encircle any nodal line, the eigenvector frame rotation becomes trivial. 

\begin{figure*}    
\centering
\includegraphics[width=0.95\textwidth]{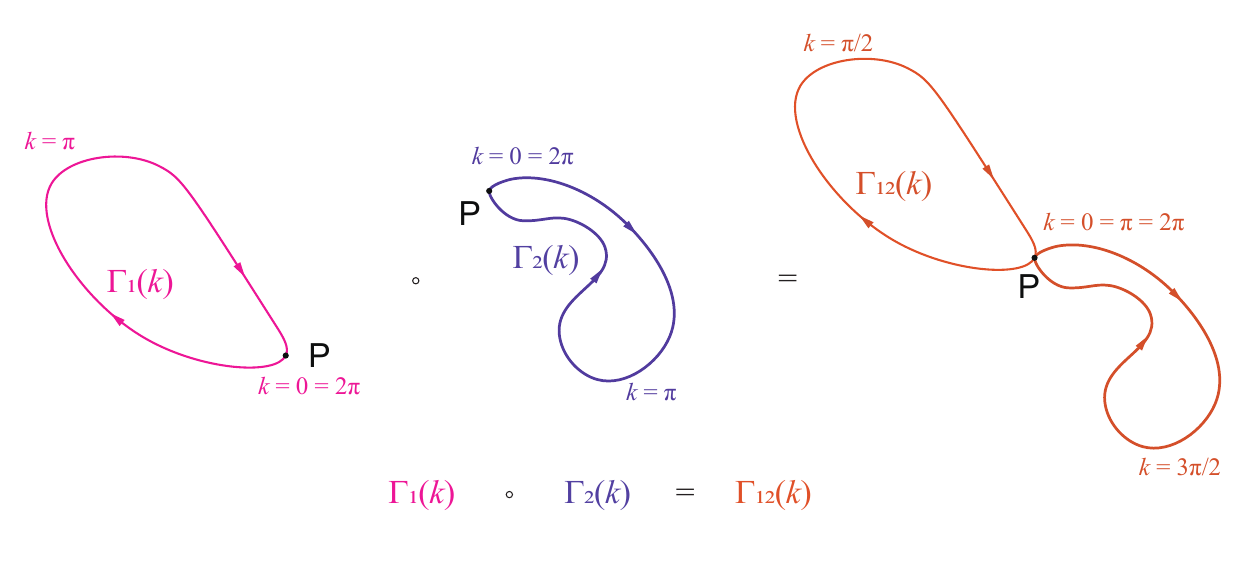}
\caption{Multiplication between loop $\Gamma_1(k)$ and loop $\Gamma_2(k)$. The product is a new loop $\Gamma_{12}(k)$.
}
\label{FIG6}
\end{figure*}

To get the dataset $V$ for clustering, we generated the closed loops set $L$ consisting of loops $\Gamma$. We set the origin of the coordinates $O(0,0,0)$ as the common base point and the initial point of all loops $\Gamma$ in $L$. 

\textbf{Sample set for Fig. \ref{FIG1}(b-f):} We generated a series of ellipses as a closed loop set $L$. These ellipses take the parameter equation:
\begin{equation}
\Gamma(k)=(\text{sin}(k),0,a(1-\text{cos(k)})), k\in[0,2\pi].
\end{equation}
Here, $a$ varies from $-\frac{10}{3}$ to $\frac{10}{3}$. In Fig. \ref{FIG6}(b), we illustrated the loops $\Gamma_1$ (colored in orange), $\Gamma_2$ (colored in light orange), $\Gamma_3$ (colored in yellow), $\Gamma_4$ (colored in green), with $a=-1.8,-1.0,1.0,1.8$, respectively.
As parameter $a$ varies continuously, the sample set intersects the blue nodal line twice (from $\Gamma_1$ to $\Gamma_2$ and from $\Gamma_3$ to $\Gamma_4$). Consequently, this sample set is divided into three segments by two phase transition points(as Fig. \ref{FIG1}(c-f) shows), yet all samples belong only to two topological phases. Samples not encircling any nodal lines belong to the trivial phase, while the samples encircling the blue nodal line, either from the upper or lower side of the origin $O$, belong to the topological phase $+i$.

\textbf{Sample set for Figs. \ref{FIG2},\ref{FIG3},\ref{FIG4}:} In order to enhance the diversity of the sample set and without loss of generality, we generated eighty loops shaped as simple closed curves (e.g., ellipses) surrounding different nodal lines. These loops belong to eight topological phases, with each phase comprising ten samples. To avoid singularities in the calculations, we omitted loops that intersect or are too close to the nodal lines. Additionally, we ensured the curves were smooth and devoid of any cusps. With these loops, we obtained the sample set $\{{H}(k)\}$ and the corresponding eigenvector frame set $\{R(k)\}$ as the input dataset for our clustering calculation.

\textbf{The product dataset for Fig. \ref{FIG3}:} 
As previously mentioned, our Hamiltonian $H(k)$ is derived from loops in the 3D $\mathbf{p}$-space. To obtain the product Hamiltonian dataset in the Multiplication Table section (the data required for constructing Fig. \ref{FIG3}), we define a ``concatenation'' operation for two loops sharing a common point, resulting in a new Hamiltonian derived from the concatenated loop (as illustrated in Fig. \ref{FIG6}. Similar to the multiplication defined for our Hamiltonian, the concatenation of two loops is simply by connecting the loops end-to-end to form a new loop. In Fig. \ref{FIG6}, the concatenation of $\Gamma_1(k)$ and $\Gamma_2(k)$ yields a new loop $\Gamma_{12}(k)$, which represents a sequence where the system first undergoes $\Gamma_1(k)$ and then $\Gamma_2(k)$. The parameter $k$ is adjusted accordingly to adapt the definition domain of the concatenated loop $\Gamma_{12}(k)$ to be within the first Brillouin zone.

\begin{figure*}[htbp]
\centering
\includegraphics[width=0.95\textwidth]{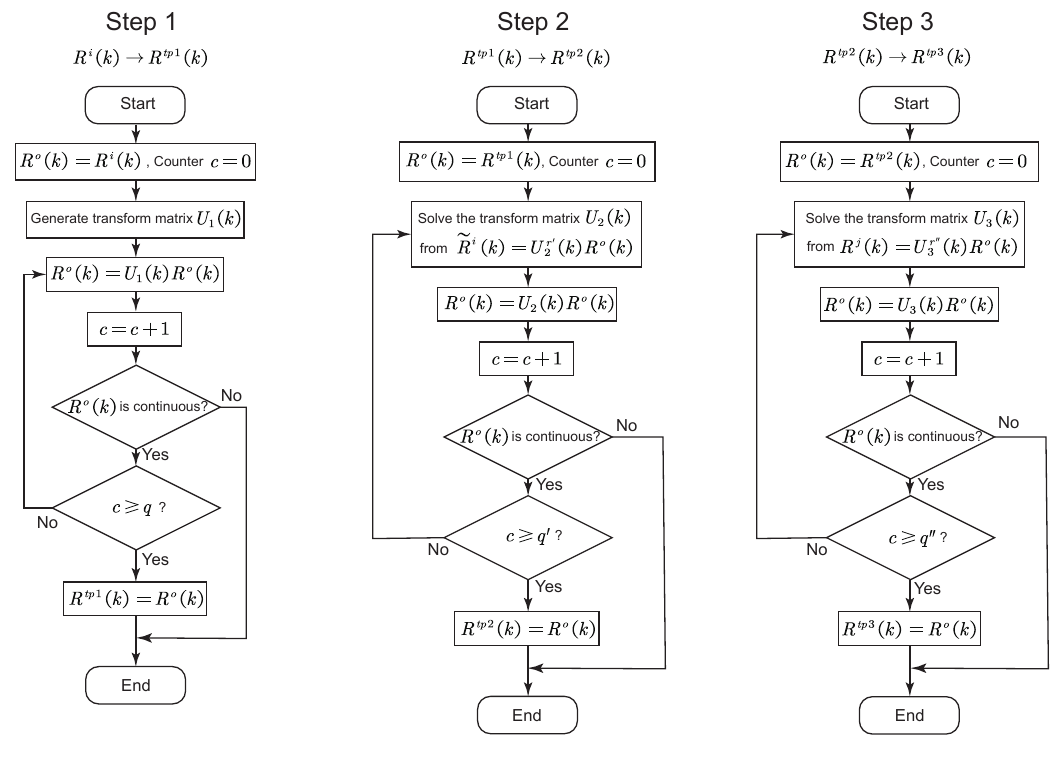}
\caption{Flowchart of Step 1, 2, 3 in the data-driven adiabatic pathfinding process.
}
\label{FIG7}
\end{figure*}

\subsection{Data-driven adiabatic pathfinding process}
In the following, we discuss the process by which our method accomplishes data-driven adiabatic pathfinding. We focus on two eigenvector frames, $R^i(k)=(|\psi^{1}_{i,k}\rangle,|\psi^{2}_{i,k}\rangle,|\psi^{3}_{i,k}\rangle)$ and $R^j(k)=(|\psi^{1}_{j,k}\rangle,|\psi^{2}_{j,k}\rangle,|\psi^{3}_{j,k}\rangle)$. $R^i(k), R^j(k)$ are not only matrices composed of eigenvectors but also belong to the special orthogonal group SO(3) and can be represented using rotation axes $\mathbf{n}_k$ and angles $\phi_k$: $R^i(k)=\text{exp}(\phi_k^i\mathbf{n}_k^i\cdot\mathbf{L}),\ R^j(k)=\text{exp}(\phi_k^j\mathbf{n}_k^j\cdot\mathbf{L})$, where $(\mathbf{L}_i)_{jk}=-\epsilon_{ijk}$ is the fully antisymmetric tensor. Our goal is to seek an adiabatic evolution path that connects the two eigenvector frames $R^i(k), R^j(k)$. In other words, we aim to identify a series of intermediate vector frames $R^o(k)$ such that the eigenvector frame $R^i(k)$ can undergo a continuous transformation through these intermediate vector frames to reach $R^j(k)$, as following:
\begin{equation}
\begin{array}{c}
	R^i(k) 
        \begin{array}{c}
	\begin{array}{c}
	\nearrow\\
	\rightarrow\\
	\searrow\\
        \end{array}
        \begin{matrix}
	R^{o1,1}(k)		\rightarrow     R^{o1,2}(k) \rightarrow \cdots\	\\
	R^{o2,1}(k)		\rightarrow 	R^{o2,2}(k) \rightarrow	\cdots\ \\	
	\vdots\ \ \ \ \ \ \ \ \   \\
	R^{oh,1}(k)		\rightarrow 	R^{oh,2}(k) \rightarrow	\cdots\ \\
        \end{matrix}
        \begin{matrix}
	R^{o1,g}(k)	\\
	R^{o2,g}(k)	\\	
	\vdots 	  \\
	R^{oh,g}(k)	\\
        \end{matrix}        
        \begin{array}{c}
	\searrow\\
	\rightarrow\\
	\nearrow\\
        \end{array}\\
        \end{array} 
        R^{j}(k)\\
\end{array}.
\end{equation}
Here, $R_{oh,g}(k)$ denotes the $g$-th intermediate vector frame in the $h$-th attempt. During the transformation process, we ensure that the symmetry of the original topological system is preserved, so the intermediate vector frame $R^o$ also satisfies $R^o(k)=(|\psi^{1}_{o,k}\rangle,|\psi^{2}_{o,k}\rangle,|\psi^{3}_{o,k}\rangle)=\text{exp}(\phi_k^o\mathbf{n}_k^o\cdot\mathbf{L})$. To ensure that the eigenvector frame undergoes a continuous transformation throughout the process, we impose continuity criteria on the intermediate states $R^o(k)$ after every transformation:
\begin{equation}\min\limits_{n,k}\left|\langle\psi_{o,k+\Delta k}^n|\psi_{o,k}^n\rangle\right|>0.7,\end{equation}
where $\Delta k=2\pi/N_k$ is the interval between our discrete $k$ values. If any intermediate vector frame does not satisfy the continuity criteria, we discard this discontinuous transformation.

We provide a detailed explanation of why the continuity criterion is necessary. When the states $|\psi_k\rangle$ is $k$ independent, $|\langle\psi_k|\psi_{k+\Delta k}\rangle|$ will take the maximum $1$. When the Hamiltonian approaches and finally reaches the phase transition point, gap closing happens and one band should exchange the wavefunctions with another band when the momentum goes from $k$ to $k+\Delta k$. It will lead to $|\langle\psi_k|\psi_{k+\Delta k}\rangle|=0 $ due to the orthogonality of wavefunctions between different bands. Due to this reason, every time we find that the intermediate state does not meet the continuity criteria, we judge that there is a phase transition during the transformation, so the transformation is not continuous anymore. In this case, we abolish this transformation.

In our method, we adopted three steps to find the adiabatic evolution path, as shown in Fig. \ref{FIG7}.

\subsubsection{Step 1: $R^i(k)\rightarrow R^{tp1}(k)$}

The purpose of Step 1 is to continuously transform the eigenvector frame $R^i(k)$ into a completely new vector frame $R^{tp1}(k)$, providing a new starting point for subsequent steps, and allowing exploration of additional adiabatic transformation paths.

Here is a detailed explanation of the flowchart for Step 1 (left panel of Fig. \ref{FIG7}). Initially, we set $R^o(k) = R^i(k)$ and initialize the loop counter $c = 0$. Next, we generate a transformation matrix $U_1(k)$ which will be applied to $R^o(k)$. Since the continuous transformation must preserve the system's symmetry, the transformation matrix $U_1(k)$ is also a rotation matrix, specifically $U_1(k) = \text{exp}(\eta\phi_k \mathbf{n}_k \cdot \mathbf{L})$. Here, $\eta$ is a parameter ensuring the transformation remains adiabatic. In our case, $\eta=0.1$. To maintain the continuity of the transformation while preserving a fixed base point, we require the rotation angle $\phi_k$ to gradually increase from zero and then decrease back to zero as $k$ varies from $0$ to $2\pi$. The magnitude of the rotation axis $\mathbf{n}_k$ must be equal to one, and to enhance the diversity, it should vary continuously. Following these guidelines, we generate a series of rotation matrices by combining some simple functions and randomly select one to serve as the transformation matrix applied to $R^o(k)$, resulting in $R^o(k) = U_1(k) R^o(k)$. After applying the transformation matrix once, we increment the loop counter $c$ by $1$. Subsequently, we assess whether $R^o(k)$ remains continuous. If $R^o(k)$ no longer meets the continuity criterion, we terminate this adiabatic pathfinding attempt. Conversely, if $R^o(k)$ satisfies the continuity criterion, we continue to apply the transformation matrix $U_1(k)$ to the updated $R^o(k)$ until the loop counter $c$ reaches or exceeds $q$. At this point, the transformation matrix $U_1(k)$ has been continuously applied to $R^i(k)$ a total of $q$ times, resulting in $R^{o}(k) = U_1(k)^q R^i(k)$. We record this transformation result as $R^{tp1}(k) = R^o(k)$, marking the outcome of Step 1. We set $q=15$ in our case.

\subsubsection{Step 2: $R^{tp1}(k)\rightarrow R^{tp2}(k)$}

The objective of Step 2 is to transform the vector frame $R^{tp1}(k)$ obtained from Step 1 into a new vector frame $R^{tp2}(k)$, bringing it closer to a fixed-axis rotation $\widetilde{R} ^{i} (k)$ through continuous transformations, thereby simplifying the form of the vector frame $R^{tp1}(k)$. This simplification can enhance the likelihood of successfully identifying the adiabatic evolution path in the subsequent step.

We will elucidate the specific process of Step 2 as depicted in the central panel of Fig. \ref{FIG7}. We initiate the procedure by setting $R^o(k) = R^{tp1}(k)$, which is the transformation result obtained in Step 1, and we initialize the loop counter $c = 0$. Subsequently, we aim to refine $R^o(k)$ so that it increasingly approximates the fixed-axis rotation $\widetilde{R}^i(k) = \text{exp}(\widetilde{\phi}_k^i \mathbf{n}_0 \cdot \mathbf{L})$. Here, $\widetilde{R}^i(k)$ is derived from the data-driven method utilizing $R^i(k)$ as a prototype for the fixed-axis rotation. To ensure that $\widetilde{R}^i(0) = \widetilde{R}^i(2\pi)$, we designate the rotation axis of $R^i(2\pi)$ as the fixed rotation axis $\mathbf{n}_0$ for $\widetilde{R}^i(k)$. For the rotation angle $\widetilde{\phi}^i_k$, for simplicity, we let it be a straightforward linear function. Thus, we obtain the fixed-axis rotation $\widetilde{R}^i(k)$. To bring the vector frame $R^o(k)$ closer to the fixed-axis rotation $\widetilde{R}^i(k)$, we solve the equation $\widetilde{R}^i(k) = U^{r'}_2(k) R^o(k)$ to determine $U_2$, where $r'$ is a sufficiently large value that ensures $U_2(k)$ represents a sufficiently small adiabatic transformation. In this context, we set $r' = 10$. We then apply the transformation matrix $U_2(k)$ to $ R^o(k)$ and update it: $R^o(k)=U_2(k)R^o(k)$. Following this transformation, the loop counter $c$ is incremented by 1. Next, we assess the continuity of $R^o(k)$; if it is not continuous, we terminate the transformation process. If $ R^o(k)$ satisfies the continuity criterion, we substitute the updated $R^o(k)$ back into the equation $\widetilde{R}^i(k) = U^{r'}_2(k) R^o(k)$ to solve for $U_2$, continuing this process until the loop counter $c\geq q'$. In our case, we set $q' = 25$. Finally, we assign the resultant $R^o(k)$ to $R^{tp2}(k)$: $R^{tp2}(k)=R^o(k)$, which serves as the outcome of Step 2.

\subsubsection{Step 3: $R^{tp2}(k)\rightarrow R^{tp3}(k)$}

The objective of Step 3 is to transform the vector frame $R^{tp2}(k)$ obtained in Step 2 into a new vector frame $R^{tp3}(k)$ through a continuous transformation, thereby bringing it as close as possible to the target vector frame $R^j(k)$. This step aims to identify an adiabatic evolution path that connects the initial eigenvector frame $R^i(k)$ to the target eigenvector frame $R^j(k)$.

The detailed process of Step 3 is illustrated in the right panel of Fig. \ref{FIG7}. The entire procedure closely resembles that of Step 2, with the distinction that the target vector frame is now $\widetilde{R}^i(k)$ instead of $R^j(k)$. Consequently, the equation for solving the transformation matrix $U_3$ is modified to $R^j(k) = U_3^{r''} R^o(k)$. Similarly, $r''$ should be large enough,  and we set $r'' = 10$. When the loop counter $c\geq q''$, we cease applying the transformation, resulting in the final vector frame $R^{tp3}(k)$. In our case, $q''=25$.

After Step 3, since $R^{tp3}(k)$ is obtained through a continuous transformation of $R^i(k)$, we substitute the local similarity $K_{tp3,j}$ between $R^{tp3}(k)$ and $R^j(k)$ for the local similarity $K_{i,j}$ between $R^i(k)$ and $R^j(k)$.

It is important to emphasize that all the steps are essential. Steps 2 and 3 directly execute transformations towards the target vector frame, which limits the ability to explore alternative adiabatic paths. Therefore, Step 1 preemptively transforms the initial eigenvector frame into various distinct vector frames, thereby providing additional opportunities for exploration. This approach is particularly crucial for the complicated dataset involving non-Abelian phases.

Through this adiabatic pathfinding process, we explore potential connections between samples, overcoming the limitations of traditional methods.

\nocite{*}
\bibliography{ref}

\end{document}